# Neural Synchronization based Secret Key Exchange over Public Channels: A survey

Sandip Chakraborty   Jiban Dalal   Bikramjit Sarkar   Debaprasad Mukherjee

Dept. of Computer Science and Engineering and Dept. of Information Technology
Dr. B. C. Roy Engineering College (West Bengal University of Technology), India

sandipch240@gmail.com   jibdalal@gmail.com   sarkar.bikramjit@gmail.com   mdebaprasad@gmail.com

*Abstract*— Exchange of secret keys over public channels based on neural synchronization using a variety of learning rules offer an appealing alternative to number theory based cryptography algorithms. Though several forms of attacks are possible on this neural protocol e.g. geometric, genetic and majority attacks, our survey finds that deterministic algorithms that synchronize with the end-point networks have high time complexity, while probabilistic and population-based algorithms have demonstrated ability to decode the key during its exchange over the public channels. Our survey also discusses queries, heuristics, erroneous information, group key exchange, synaptic depths, etc, that have been proposed to increase the time complexity of algorithmic interception or decoding of the key during exchange. The Tree Parity Machine and its variants, neural networks with tree topologies incorporating parity checking of state bits, appear to be one of the most secure and stable models of the end-point networks. Our survey also mentions some noteworthy studies on neural networks applied to other necessary aspects of cryptography. We conclude that discovery of neural architectures with very high synchronization speed, and designing the encoding and entropy of the information exchanged during mutual learning, and design of extremely sensitive chaotic maps for transformation of synchronized states of the networks to chaotic encryption keys, are the primary issues in this field.

*IndexTerms*—Cryptography, Key exchange, Neural networks, Synchronization.

## I. INTRODUCTION

Cryptography is the practice of constructing and analyzing protocols for secure exchange of information i.e. communication, overcoming the presence or influence of adversaries or third parties, e.g. preventing leakage of information to unauthorized parties. It deals with various aspects in information security e.g. data confidentiality, data integrity, and authentication. Cryptographic algorithms are designed around computational hardness assumptions e.g. using number-theoretic concepts, making such algorithms hard to break in practice by any unauthorized party. These algorithms/schemes are termed as computationally secure since it is infeasible to break them by known practical means, although they are breakable, in theory. Theoretical advances e.g. in integer factorization algorithms, and faster computing technology require these schemes to be continually improved [1]. In 1970s, Diffie & Hellmann found that a common secret key could be created over a public channel accessible to any unauthorized party [2]. Since then, many public key cryptosystems have been presented which are based on number theory, and they demand large computational power [3]. Moreover the processes involved in generating public key are very complex and time consuming. To overcome these disadvantages, several other concepts and techniques have been explored. Among them, it has been found that the concept of neural networks can be used to generate common secret key, and this can offer one of several possible solutions to this critical issue of key exchange.

The paper is organized as follows. We introduce the concepts of synchronization and chaos in artificial neural networks in Section II. In Section III we discuss the basic model of neural synchronization based key exchange and in Section IV we survey the other basic models. In Section V we discuss the different types of attacks, in Section VI the parity machines, in Section VII Queries, Synaptic Depth and Erroneous Information, in Section VIII we survey the most important studies in recent times. Finally in Section IX some additional forms of cryptography and their corresponding noteworthy studies have been mentioned. Section X summarizes and concludes the survey.

## II. NEURAL NETWORKS, SYNCHRONIZATION AND CHAOS

It is widely known that artificial neural networks are computational models inspired by animal brains that are capable of machine learning and pattern recognition. Neural networks can learn from a training data set and then can predict or classify data. They are usually presented as systems of interconnected neurons that can compute outputs from inputs by propagating information through the network. Neural networks are used as a form of soft computing technique to solve computationally difficult problems e.g. speech and face recognition, gene prediction etc. [4]. These networks constitute a form of dynamic systems which show a variety of complex behaviour including the phenomenon of chaos and synchronization [5]. Chaotic systems are nonlinear dynamical systems that are highly sensitive to initial conditions.

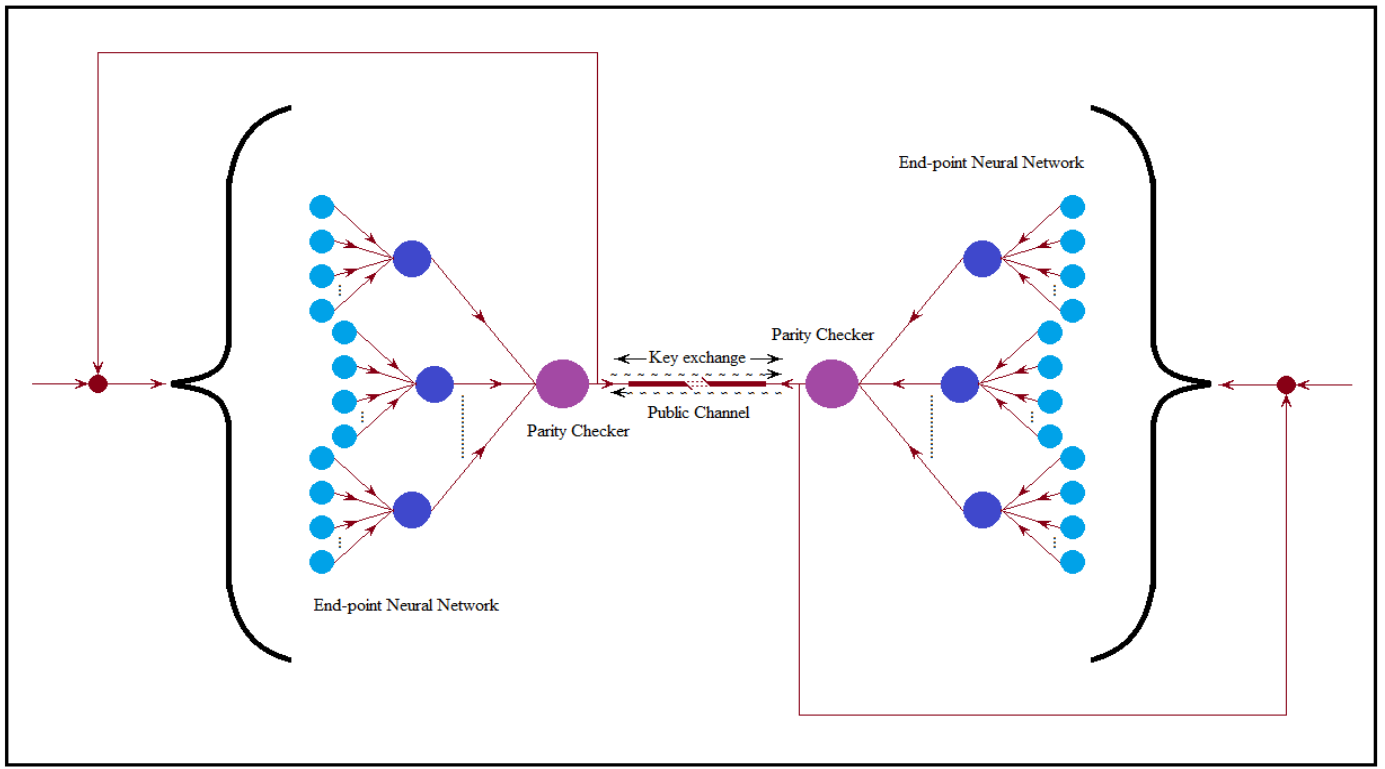

Fig. 1. Schematic of a public channel using neural synchronization based key exchange

In these systems, minute differences in initial conditions generate highly diverging outputs, making long-term prediction very difficult in general. This happens even though these systems are deterministic, meaning this kind of behaviour exist even when their future dynamics is fully determined by their initial conditions, with no random elements involved. One popular model of chaotic systems is chaotic maps [6]. Chaotic maps are mathematical evolution functions which show chaotic behaviour, and may be parameterized by a discrete time or continuous time parameter. Chaotic systems sometimes show the property of synchronization. Synchronization is the coordination of events to operate a system in unison, or, the attainment of equivalent states in systems while interacting with each other. Most systems esp. the stochastic ones may be only approximately synchronized. Phase synchronization, one of the widely used concepts of synchronization, is the process by which two or more cyclic signals tend to oscillate with a repeating sequence of relative phase angles [5]. One form of synchronization which is already known in encryption systems is the validation that the receiving ciphers are decoding the right bits at the right time.

Thus, it seemed that using synchronization and chaotic characteristics of dynamical systems may provide a promising direction to the design of new and efficient cryptosystems. Following this promise, in recent years, artificial neural networks have been applied and experimented with, in several forms of cryptographic systems/techniques.

### III. FUNDAMENTAL MODEL

In 2002, Kanter et. al. published a series of related papers where they have shown the beautiful connection between neural networks and cryptography [7, 8, 9]. They demonstrated that synchronization of neural networks can lead to a method of exchange of secret messages or keys. They demonstrated that when two artificial neural networks are trained by suitable learning rules e.g. Hebbian rules, on their mutual outputs, then these networks can develop equivalent states of their internal synaptic weights, i.e., the networks synchronize to a state with identical time dependent weights. These synchronized weights are then used to construct a key exchange protocol. They found that it was impossible to decrypt the secret message even for an opponent who knew the protocol and all details of any transmission of the data. They also proved that this was primarily because the tracking of the weights of neural networks during synchronization was a NP- hard problem. But, on the other hand, the complexity of the generation of the secure channel is linear with the size of the network. These results opened up new avenues in modern cryptography, and showed how synchronization by mutual learning in neural networks can be applied to secret key exchanges over public channels. This and similar results later gave rise to the field of Neural Cryptography. Thereafter, researchers have tried several different methods for cryptography using neural networks in various forms. (Fig.1.)

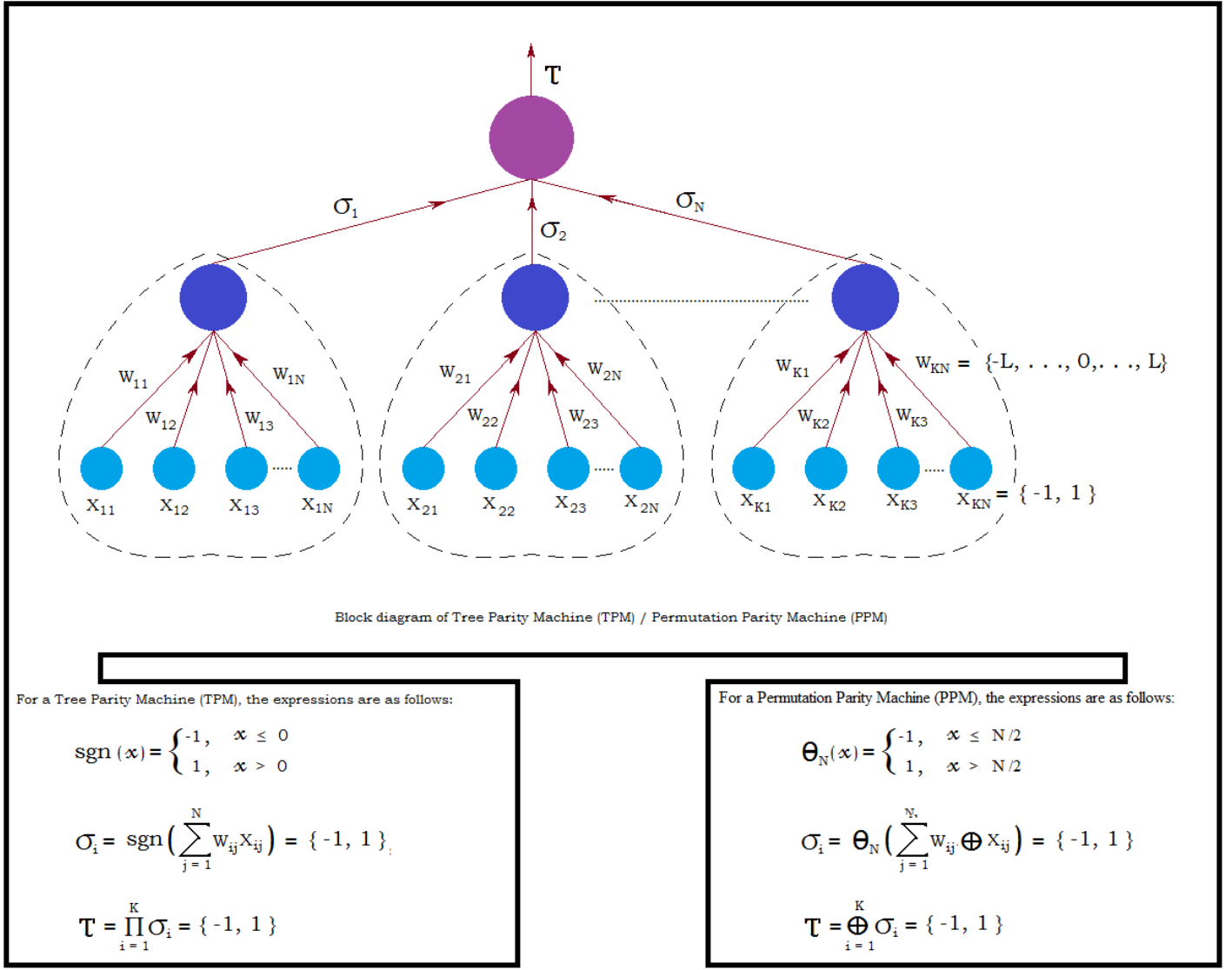

Fig. 2. Schemata and models of Tree Parity Machine (TPM) and Permutation Parity Machine (PPM)

Here, in this paper, we survey and discuss some of the most prominent problems addressed and some of their solutions in the domain of neural cryptography.

Neural cryptography deals with the problem of key exchange between two neural networks using the mutual learning concept. Two neural networks that are trained on their mutual output synchronize to an identical time dependant weight vector. The two networks exchange their outputs (in bits) and the key between the two communicating parties is eventually represented in the final learned weights, when the two networks are said to be synchronized. This novel phenomenon has been used for creation of secure cryptographic secret-keys on a public channel. In this form of neural synchronization based key exchange, the security of the process is put at risk if an external attacker is capable of synchronizing with any of the two parties i.e. the end point neural networks, during the learning (or training) process.

## IV. OTHER BASIC MODELS

In 2002, Kinzel and Kanter further analyzed their previous model (discussed above) using multilayer neural networks with discrete weight [10]. They found further validation of their previous results i.e. the in this case with discrete weights too, the networks synchronized with identical final internal weights. The authors also further validated the primary assumption, that synchronization by mutual learning between neural networks can be applied to generate secret public channel keys, through theoretical justification. In the same year, Rozen et.al. studied the mutual learning process between two parity feed-forward neural networks [7]. They designed these networks with both discrete and continuous weights. It was found that only finite steps were required to attain full synchronization between the two networks having discrete weights. The learning time of an attacker that is trying to imitate one of the networks was found to be much longer than the synchronization time of the networks. This particular finding strengthened the possibility of overcoming the influence of attackers or adversaries, because the result indicated that the keys can be exchanged before the attacker can decode the them (the keys). In 2003, Milsovaty et al. developed the concept of two step synchronization for generation of secret encryption keys over public channels [11]. In the first step, the external signal to the system is firstsynchronized by neural networks. The neural networks synchronize amongst themselves by the process of mutual learning, as explained previously. Thereafter, the common synchronized signal is applied as input to a set of chaotic maps (defined previously). These chaotic maps then generate the secret encryption keys to be used over public channels. This makes the network or process for generation of the keys a hybrid one- i.e. a hybrid between neural networks and chaotic maps. Thesecurity of key exchange over public channels was shown to be increased by this form of chaotic synchronization.

## V. TYPES OF ATTACKS

Public channels exchanging secret keys through neural synchronization may experience primarily four kinds of attacks: simple, geometric, majority and genetic [12]. In simple attack, the attacker's neural network has the same structure as that of the end point networks. All that the attackers have to do is to start with random initial weights and to train with the same inputs transmitted between the end point networks. A geometric attack, on the other hand, can outperform a simple attack, because, in addition to applying the same learning process, the attacker can utilize the output of the attacker machine and the local fields of its hidden units. The majority attack or the cooperative attack is similar to the geometric attack, but here the attacker can increase the probability to predict the internal representation of any of the partners' neural network, by using several networks, with random weight vectors, working in a cooperating group rather than as individuals. In the genetic attack, the attacker starts with only one network but is permitted to use more networks as synchronization progresses. In 2004, Shacham et al. have shown that a group of attackers cooperating amongst themselves (i.e. cooperative attack) throughout the synchronization process can have high degrees of success in breaking the neural synchronization based key exchange [13].

## VI. PARITY MACHINES

One of the most popular models of neural networks i.e. feed forward networks as the end points of the communicating channels is the tree parity machine (TPM). The TPM is a special type of multi-layer feed-forward neural network [14].

It consists of one output neuron, K hidden neurons and K*N input neurons. Inputs to the network are binary: $X_{KN} = \{-1, 1\}$

The weights between input and hidden neurons take the values: $W_{KN} = \{-L, \ldots, 0, \ldots, L\}$

Output value of each hidden neuron is calculated as a sum of all multiplications of input neurons and these weights:

$$\sigma_i = \text{sgn}\left(\sum_{j=1}^{N} W_{ij} X_{ij}\right) = \{-1, 1\}$$

Signum is a simple function, which returns -1,0 or 1:

$$\text{sgn}(x) = \begin{cases} -1, & x \leq 0 \\ 1, & x > 0 \end{cases}$$

If the scalar product is 0, the output of the hidden neuron is mapped to -1 in order to ensure a binary output value.

The output of neural network is then computed as the multiplication of all values produced by hidden elements:

$$\tau = \prod_{i=1}^{K} \sigma_i = \{-1, 1\}$$

Output of the tree parity machine is binary. (Fig.2.)

Another widely used model of such end point networks is a binary variation of the TPM- the Permutation Parity Machine [15]. Another upcoming model of such networks, which is also a variation of the Tree Parity Machine, is the c, which has almost similar topology to the TPM [16].

## VII. QUERIES, SYNAPTIC DEPTH AND ERRONEOUS INFORMATION

In 2005, Ruttor et al. extended previous key-exchange protocols in a novel way. They introduced the concept of querie [17]. More information, in the form of queries, was used to train the neural networks at the two end points of the channel. Random inputs to these networks were replaced with queries, which were a function of the current state of the neural networks. The authors demonstrated that these queries improved the security against cooperating or majority attacks. This was hailed as a major result in the field. As previously Shacham et al. showed that cooperative attackers can break the security provided by neural synchronization based cryptography. Furthermore, these authors had also demonstrated that the success probability of the attacker, even if it is a cooperative attacker, can be reduced without increasing the average synchronization time of the neural networks. The concept of queries was further extended and explored later. It was demonstrated that if the connections (end points of the channels) are generating inputs which are correlated with their state, and the end points are asking their partner for the corresponding output bit, then the overlap between input and weight vector is so low that the additional information does not reveal much about the internal states [18]. In this way the queries introduced a mutual influence between end points which was not available to the attacking network. When queries are incorporated along with the Hebbian training rule, then it was found that the probability of successful attack decreases significantly, even for cooperative attacks [19].

Again in 2006, Ruttor et al. demonstrated and also proved that incrementing the synaptic depth of the neural networks increased the synchronization time of the networks, but only polynomially [20]. But this ensured that the probability of successful geometric attacks on the neural synchronization based key exchange system is reduced exponentially. They utilized the genetic algorithm method to select the fittest neural networks for such a process. Furthermore, it was shown in their work that the number of networks needed for a majority/cooperative attack to be successful grows exponentially with increasing synaptic depth.

As discussed previously, security of neural synchronization based key exchange is put at risk if an attacker is capable of synchronizing with any of the two parties during the training or learning process. Therefore, diminishing the probability of such a threat improves the reliability of exchanging the output bits through a public channel. The synchronization with feedback algorithm is one of the existing algorithms that enhance the security of neural cryptography [21]. In the beginning of this decade (2010), algorithms were proposed to enhance the mutual learning process. They mainly depended on disrupting the attacker confidence in the exchanged outputs and input patterns during training. One such algorithm relies on one party sending erroneous output bits, with the other party being capable of predicting and correcting this error [12]. Another such proposed algorithm functions in a way where inputs are kept partially secret and the attacker has to train its network on input patterns that are different from the training sets used by the communicating parties. It was also shown that a hybrid of these two approaches is also satisfactory.

## VIII. OTHER RECENT STUDIES

Side channel attacks on neural synchronization models of key exchange have also been evaluated. In a paper during the same time when the above algorithms got published, the authors investigated schemes of side channel attacks and their effects on communicating channels with tree parity machine models of neural networks at both end points [22].

To test the validity of the neural synchronization based key exchange algorithms on models other than the TPM, a key exchange algorithm was also proposed based on multilayer feed-forward neural networks incorporating permutation parity machines [23]. This algorithm's performance against the common attacks on such channels e.g. simple, geometric, majority and genetic was exhaustively studied and the performance (based on various metrics) was found quite satisfactory. Other proposals also tested the utility of the PPMs for key exchange [15, 24].

Neural cryptography using a form of information substitution was also proposed in 2011 [25]. In this approach- recursive positional modulo-2 substitution, both the communicating networks received an indistinguishable input vector, and produced an output bit. Then these networks were trained based on these output bits. Based on this scheme, a secret-key, of variable length, was formed. The original information, which was represented in plain text format, was encrypted using the substitution technique, i.e. recursive positional modulo-2 substitution. The intermediate-level cipher information was again encrypted to form the final cipher text. This was done through chaining and cascaded *xor*ing of identical weight vectors with the identical length intermediate cipher text block. The receiver then used identical weight vector for performing deciphering process for getting the encrypted cipher text and secret key for decoding. This particular method proved to be an inspiration for some later exploratory work on neural synchronization based variable length keys for encrypting symbolic information. In addition to this, an important study has also been performed for group key exchange schemes. Here, a group of end point parties can share a common key. A recursive algorithm was proposed, where it was shown that a group of N parties is able to share a common key in an acceptable time (sup ($\log_2(N)$)) [26]. This acceptable time complexity primarily depended on the topological arrangement of the parties in the group withrespect to each other. In this particular work, the end point communicating parties were ordered on variations of binary tree topologies, e.g. swapping and election.

But in 2012, an algorithm was proposed which demonstrated that successful attacks are quite possible on permutation-parity-machine-based neural cryptosystems. This particular algorithm executed a probabilistic form of an attack on the key-exchange protocol [15]. Previously such algorithms/schemes of attacks that worked by imitating the synchronization of the communicating partners had been proposed (discussed previously in this paper). But here, a Monte Carlo method was used to sample the space of possible weights during inner rounds of learning. Thereafter, the sampled inner round information was conveyed from one outer round of learning to the next one. It was shown in this work that the mutual learning based protocol failed to synchronize faster than an attacker who used this sampling based inner and outer round learning. This paper has been recognized as opening up of new areas of investigation in neural cryptography, esp. for designing faster synchronizing protocols between the end point neural networks. In the current year (2013), another work also analysed the utility and degree of success of chaotic and Sequential Machine models of neural network synchronization [27].

As research in this field progressed, a general consensus started emerging that, till now, the tree parity machine (TPM) network with hidden unit K=3 is the model that is most suitable for a neural protocol [28]. Thus, it becomes important to find more variety of neural network architectures and synchronizing mechanisms that provide enhanced security to different forms of attacks.

Thus, in 2013, a two-layer tree-connected feed-forward neural network (TTFNN) model was proposed. This model utilized the concept that two communicating partners are capable of exchanging a vector with multiple bits in each time step [28]. In this work, feasible conditions and heuristic rules that would make the neural synchronization based protocol successful against common attacks were obtained. The authors proposed methods for developing TTFNN based schemes, and discovered cases of that have better synchronization speed than some of the previously published protocols based on the established TPM model of the communicating partners.

In another analysis, the security of neural cryptography systems was studied by investigating the dynamics of information leakage through the learning process [29]. This study also analysed how the leaked information can be used to reduce the complexity of common attack strategies. In a different but seminal study very recently, a generalized architecture of the end point neural networks, actually extended and unified models of tree parity machine (TPM) and tree committee machine (TCM), were proposed [30]. In addition to this, the authors also proposed a heuristic rule for such generalized architectures. The algorithm incorporating the generalized architectures seemed to have better performance than the protocols using the Tree Parity Machine models of the end point networks (like some of the previously discussed ones). In this proposed generic framework the authors also found that the heuristic rule can actually improve the security to common attacks to their scheme.

Thus, we have discussed and surveyed the most important contributions in the recent past on neural synchronization based key exchange, and also the algorithms and results that have defined this field.

## IX. ADDITIONAL FORMS OF NEURAL CRYPTOGRAPHY

In addition to the core issues of neural cryptography that we have discussed in this paper, neural networks have put to good use for other important issues in general cryptography e.g. generation of cipher text, image encryption, generation of image keys, generation of pseudo random numbers, construction of well-formed has functions for transformation of messages, etc. For algorithms for such purposes, neural models incorporating delay, chaos & hyper-chaos, state machines, switching, dimension reduction, Hopfield, Hebbian and random walk dynamics, and other similar properties. Out of these, chaotic and delay based neural models have high visibility in research literature [31, 32].

Previous to the form of neural synchronization based cryptography, a form of cryptography, called visual neural cryptography became popular (and still is), when Yue and Chiang in 2000 proposed a neural network based approach for the encryption of images [33]. The neural networks used here took gray level images as input and processed them to produce binary images in the output. Since then, it has been found that these neural net based models for image encryption has been able to address issues with various access schemes, and complex images. In a widely acknowledge work in 2009, Penget. al. proposed a digital image encryption algorithm using novel model based on hyper-chaotic cellular neural networks [34]. The authors performed key space analysis, sensitivity analysis, information entropy analysis and correlation coefficients analysis of adjacent pixels for the algorithm, and found its results quite satisfactory. In another parallel development, some researchers in neural cryptography have proposed novel models using neural networks for generating cipher text. Among them, one of the most influential models is that of Hopfield neural networks with embedded chaotic behaviour. In 2006, Yu and Cao proposed such a model with time varying delay [35]. The chaotic neural network was primarily used for generation of binary strings or sequences. These binary sequences were used for masking plain text. The method they proposed for masking the text was to switch off the neural network maps, and subsequent permutation of the generated binary sequences. Several papers have validated this idea in recent times and have proposed this model as a significant methodology for secure transmission of large complex multidimensional data [36]. For a different kind of issue in cryptography, recently, in 2009, Xiao et. al. proposed a novel scheme to overcome issues of hash functions in parallel computing models [37]. They proposed an algorithm for the construction of parallel keyed hash functions. They used the chaotic neural network model to study such hash functions. They developed two mechanisms, one was of changeable-parameter and the other was of self-synchronization. These two mechanisms established a well-formed relation between the hash value bits and the message. The authors claimed that the algorithm structure ensured the uniform sensitivity of the hash value to the message blocks at different positions. Although this claim has been disputed later on by several researchers, but their basic model has found validation in some other popular schemes [38].

## X. SUMMARY AND CONCLUSION

Cryptographic algorithms are primarily based upon number theoretic concepts, e.g. integer factorization, which has high time-complexity. Neural cryptography, and particularly neural synchronization using learning rules e.g. Hebbian, random walk, based exchange of secret keys over public channels, offer a novel, appealing and useful alternative. The concept originated in a mature form in the beginning of this millennium through the work of Kinzel and Kanter. They showed that two neural networks undergoing the process of mutual learning can synchronize to a symmetric internal state, and this state can be used as a key over public channels. Several forms of attacks are possible on this scheme, but deterministic algorithms to synchronize with the end-point networks have high time complexity, while probabilistic and population-based algorithms have demonstrated ability to decode the public key during its exchange over the channels. But, schemes have also been proposed to increase the time complexity of any algorithm used to intercept the key while it is being exchanged, e.g. by using queries, heuristics, error bits, group key exchange, longer synaptic depths, etc. The Tree Parity Machine and its variants - neural networks with tree topologies incorporating parity checking of state bits, have been found to be one of the most secure models of the end-point networks. One of the primary challenges in this field of neural cryptography appears to be the discovery of neural architectures with very high synchronization speed, and designing the encoding & entropy of the information exchanged during mutual learning, to prevent the synchronization of an attacker during the mutual learning process. A promising future direction may be the study of chaotic maps for transformation of synchronized states of the networks to chaotic encryption keys, with exceptionally low tolerance for decryption error. In addition to all the above, neural networks have also been used for other aspects of cryptography e.g. text and image encryption, generation of pseudo-random numbers for text and image keys, construction of cryptographic hash functions, etc.